\documentstyle[12pt,a4]{article}

\def\L{ {\cal L}}

\def\mxth{\mathsurround=0pt }
\def\xversim#1#2{\lower2.pt\vbox{\baselineskip0pt \lineskip-.5pt
x  \ialign{$\mxth#1\hfil##\hfil$\crcr#2\crcr\sim\crcr}}}

\def\slash{\llap /}
\def\lagr{{\cal L}}
\newcommand{\thd}{{\theta_{\scriptscriptstyle\rm D}}}
\newcommand{\kild}[1]{{k^{#1}_{\scriptscriptstyle\rm D}}}

\renewcommand{\a}{\alpha}
\renewcommand{\b}{\beta}

\renewcommand{\d}{\delta}

\newcommand{\pa}{\partial}
\newcommand{\g}{\gamma}
\newcommand{\G}{\Gamma}

\newcommand{\e}{\epsilon}
\newcommand{\z}{\zeta}

\renewcommand{\L}{\Lambda}
\newcommand{\m}{\mu}

\newcommand{\s}{\sigma}

\newcommand{\Ka}{{K\"ahler}}
\renewcommand{\O}{{\Omega}}

\def\be{\begin{equation}}
\def\ee{\end{equation}}
\def\bea{\begin{eqnarray}}
\def\eea{\end{eqnarray}}

\newcommand{\ft}[2]{{\textstyle\frac{#1}{#2}}}

\newcommand{\eqn}[1]{(\ref{#1})}

\begin{document}
\begin{titlepage}
\begin{center}
\hfill THU-98/30  \\[2mm]
\hfill {\tt hep-th/9808160}\\
\vskip 12mm

{\Large {\bf Rigid $N\!=\!2$ superconformal hypermultiplets}}\footnote{%
  Talk given by B. de Wit at the International Seminar ``Supersymmetries 
  and Quantum Symmetries", July 1997, Dubna. The 
  content of this contribution is related to the actual 
  presentation at the meeting, but takes into account more recent 
  developments.}

\vskip 10mm

\bf{Bernard de Wit$^a$, Bas Kleijn$^a$ and 
Stefan Vandoren$^b$}

\vskip 8mm

$^a${\em Institute for Theoretical Physics, 
Utrecht University,}\\
{\em 3508 TA Utrecht, Netherlands}\\
{\tt  B.deWit@phys.uu.nl, B.Kleijn@phys.uu.nl} \\[2mm]
$^b${\em Department of Physics, University of Wales Swansea,}\\
{\em  SA2 8PP Swansea, U.K.} \\
{\tt pysv@swansea.ac.uk}

\vskip 8mm

\end{center}

\vskip .2in

\begin{center} {\bf ABSTRACT } \end{center}
\begin{quotation}\noindent 
We discuss superconformally invariant systems of
hypermultiplets coupled to gauge fields associated with
target-space isometries. 
\end{quotation}

\vfill

August 1998\\

\end{titlepage}

\eject

\section{Introduction}
Hypermultiplets played an important role in the work of 
Victor~I.~Ogievetsky, to whose memory this meeting is dedicated. 
We remember him as a devoted scientist, but above all as a dear 
friend and colleague who is sorely missed. 

In this contribution we discuss hypermultiplets coupled to gauge  
fields whose action is invariant under rigid $N=2$ superconformal 
symmetries. This study is both motivated by recent interest in 
superconformal theories \cite{sconf}  and by our attempts 
to understand the coupling of hypermultiplets to supergravity in 
a way that is more in parallel with the special-geometry 
formulation of vector multiplets \cite{DDKV}. In this respect it 
is important that we employ an on-shell treatment of 
hypermultiplets (to avoid  
an infinite number of fields), while the vector 
multiplets and the superconformal theory are considered fully 
off-shell. This implies that the algebra of the superconformal 
and gauge symmetries is known up to the hypermultiplet field 
equations.  

\section{Hypermultiplet Lagrangians}
\setcounter{equation}{0}

Hyper-\Ka\ spaces serve as target spaces for nonlinear sigma
models based on hypermultiplets \cite{BagWit}. We start here by
summarizing some results on the formulation of these theories 
following \cite{DDKV}. With respect to the results of 
\cite{BagWit} this 
formulation differs in that it incorporates both a metric
$g_{AB}$ for the hyper-K\"ahler target space and a metric
$G_{\bar \a\b}$ for 
the fermions. Here we assume that the $n$ hypermultiplets are
described by $4n$ real scalars $\phi^A$, $2n$ positive-chirality  
spinors $\zeta^{\bar \a}$ and $2n$ negative-chirality spinors 
$\zeta^\a$. The latter two are related by complex conjugation (so
that we have $2n$ Majorana spinors) under which 
indices are converted according to $\a\leftrightarrow \bar \a$,
while SU(2) indices $i,j, \ldots$ are raised and lowered. The
presence of the fermionic metric is important in obtaining the
correct transformation rules under symplectic transformations
induced by the so-called $\bf c$-map from the electric-magnetic
duality transformations on a corresponding theory of vector
multiplets. In formulations based on $N=1$ superfields (such as
in \cite{HKLR}) one naturally has a fermionic metric but of a special
form. 

The supersymmetry transformations are parametrized in terms of 
certain $\phi$-dependent quantities $\g^A$ and $V_A$ as
\be
\begin{array}{l}
\d_{\scriptscriptstyle\rm Q}\phi^A= 2( \g^A_{i\bar\a} \,\bar\e^i 
\zeta^{\bar \a} +  
\bar\g^{Ai}_{\a} \,\bar\e_i \zeta^\a )\,,
\end{array}
\qquad
\begin{array}{l}
\d_{\scriptscriptstyle\rm Q}\zeta^\a  =  V_{A\,i}^\a \,\pa\slash\phi^A\e^i 
-\d_{\scriptscriptstyle\rm Q}\phi^A\, \G_{A}{}^{\!\a}{}_{\!\b}\,\zeta^\b \,, \\[1mm]
\d_{\scriptscriptstyle\rm Q}\zeta^{\bar \a}= \bar V_A^{i\bar\a} \,\pa\slash\phi^A\e_i 
-\d_{\scriptscriptstyle\rm Q}\phi^A\,\bar\G_{A}{}^{\!\bar\a}{}_{\!\bar \b} \,\zeta^{\bar \b} 
\,. 
\end{array} \label{4dhsusy}
\ee
Observe that these variations are consistent with a U(1) chiral
invariance  
under which the scalars remain invariant, which we will denote by 
${\rm U(1)}_{\rm R}$ to indicate that it is a subgroup  
of the automorphism group of the supersymmetry algebra. In
section~4 this U(1) will be included as one of the conformal
gauge groups. However, 
for generic $\g^A$ and $V_A$, the SU(2)$_{\rm R}$ part of the 
automorphism group cannot be realized consistently on the fields. 
In the above, we only used that $\z^\a$ and $\z^{\bar \a}$ are related 
by complex conjugation. 

The Lagrangian takes the following form
\be
 \lagr= -\ft12 g_{AB}\,\pa_\m\phi^A\pa^\m\phi^B  
-G_{\bar \a \b}( \bar\zeta^{\bar \a} D\!\slash \,\zeta^\b +  
\bar\zeta^\b D\!\slash \,\zeta^{\bar\a}) -\ft14  W_{\bar
\a\b\bar\g\d}\, \bar \zeta^{\bar \a}  
\g_\m\zeta^{\b}\,\bar \zeta^{\bar \g} \g^\m\zeta^\d 
\,, \label{4dhlagr1}
\ee
where we use the covariant derivatives
\be
D_\m \zeta^\a= \pa_\m \zeta^\a + \pa_\m\phi^A\, 
\G_{A}{}^{\!\a}{}_{\!\b} \,\zeta^\b\,, 
\quad
D_\m \zeta^{\bar\a}= \pa_\m \zeta^{\bar \a}  +\pa_\m\phi^A\,\bar 
\G_{A}{}^{\!\bar\a}{}_{\bar \b} \,\zeta^{\bar \b} \,.
\ee
Besides the Riemann curvature $R_{ABCD}$ we will be dealing with
another curvature $R_{AB}{}^{\!\a}{}_{ \b}$ associated with the
connections $\G_{A}{}^{\!\a}{}_{\!\b}$, which takes its values in
$sp(n)\cong usp(2n;{\bf C})$. The tensor $W$ is defined by  
\be
W_{\bar \a \b \bar \g\d} = 
R_{AB}{}^{\!\bar\e}{}_{\bar \g} \,\g^A_{i\bar\a}\,\bar \g^{iB}_\b\, 
 G_{\bar \e\d}  = \ft12 R_{ABCD} \,\g^A_{i\bar\a}\,\bar \g^{iB}_\b\, 
\g^C_{j\bar\g} \,\bar \g^{jD}_\d\, . \label{defW}
\ee
Most of these quantities are not independent, as we shall specify
below, and the models are entirely characterized by the
target-space geometry (for instance, encoded in the metric
$g_{AB}$) and 
the Sp($n$)$\times$Sp(1) one-forms $V^\a_i= V^\a_{A \,i}\,{\rm
d}\phi^A$. The Sp(1) factor is associated with the indices $i,j,
\ldots$, and coincides with the SU(2)$_{\rm R}$ group mentioned 
above.   

The metric $g_{AB}$, the tensors
$\g^A$, $V_A$ and the fermionic metric $G_{\bar\a\b}$ are all
covariantly constant with respect to the Christoffel connection
and the connections $\G_{A}{}^{\!\a}{}_{\!\b}$. Furthermore
we note the following relations,
\bea
&&\g^A_{i\bar\a} \,\bar V_B^{j\bar\a} + \bar \g^{A\,j}_\a \,V_{B\,
i}^\a  = \d^j_i\,\d^A_B\,, \label{clifford}  \nonumber \\
&&g_{AB}\, \g^B_{i\bar \a} = G_{\bar\a\b}\, V_{A\,i}^\b 
\,,   \qquad
\bar V^{i\bar \a}_A \, \g^A_{j\bar \b} = \d^i_j\, 
\d^{\bar \a}_{\,\bar \b}\,. \label{inverse}
\eea
These conditions define a number of useful relations between
bilinears\footnote{%
  Such as
  $
  \bar \g_{A\a}^j \,V_{Bi}^\a = \gamma_{Bi\bar\alpha}
  \,\bar V^{j\bar \a}_A = - \bar \gamma^j_{B\alpha}
  \,V^\a_{i A} + \d^j_i \,g_{AB}$.} %
which include three antisymmetric covariantly constant 
target-space tensors, 
\be
J^{ij}_{AB} = \g_{Ak\bar\a}\,\varepsilon^{k(i}\bar V^{j)\bar\a}_B\,,
\ee
that span the complex structures of the
hyper-K\"ahler target space. They satisfy 
\be
(J^{ij}_{AB})^\ast =
\varepsilon_{ik}\varepsilon_{jl}\,J^{kl}_{AB}\,,\qquad 
J^{ij C}_{\,A} \,J^{kl}_{CB} = \ft12
\varepsilon^{i(k}\varepsilon^{l)j}\, g_{AB}
+ \varepsilon_{~}^{(i(k} \, J^{l)j)}_{AB}\,.
\ee
In addition we note the following useful identities, 
\be
\g_{Ai\bar\a}\, \bar V^{j\bar \a}_B= \varepsilon_{ik} J^{kj}_{AB} +
\ft12 g_{AB}\, \d^j_i\,,\qquad J_{AB} ^{ij}\, \gamma_{\bar \a
k}^B= -\d^{(i}_k \varepsilon_{~}^{j)l}\,\gamma_{Al\bar\a}\,. 
\ee
We also note the existence of covariantly constant 
antisymmetric tensors,  
\be
\O_{\bar\a\bar\b} =\ft12  \varepsilon^{ij}\,g_{AB}\, \g^A_{i\bar 
\a}\,\g^B_{j\bar \b}\,,
\quad 
\bar\O^{\bar\a\bar \b} =\ft12  \varepsilon_{ij}\,g^{AB}\, \bar 
V_A^{i\bar \a}\,\bar V_B^{j\bar \b}\,,
\ee
satisfying $\O_{\bar\a\bar\g}\,\bar \O^{\bar\g\bar\b} = -\d^{\bar \b}_{\bar 
\a}$. 

The existence of the covariantly constant tensors implies a variety 
of integrability conditions for the curvature tensors.
For instance, one proves that the Riemann curvature and the
Sp$(n)$ curvature are related, as indicated in  
\eqn{defW}. 
The tensor $W$ defined in \eqn{defW} can also be written as
$W_{\a\b\g\d}$ by contracting  
with the metric $G$ and the antisymmetric tensor $\O$. It then 
follows that $W_{\a\b\g\d}$ is symmetric in symmetric index pairs 
$(\a\b)$ and $(\g\d)$. Using the Bianchi identity for Riemann 
curvature, which implies $g_{D[A}R_{BC]}{}^{\!\bar \b}{}_{\bar\a} 
 \,\g^D_{i\bar\b} = 0$, one shows that it is in fact symmetric in 
all four indices. 
For further results and discussion we refer to \cite{DDKV}. In 
the next section we consider the gauging of invariances of the 
hypermultiplet action. Such invariances are related to isometries 
of the hyper-K\"ahler manifold. These isometries have been 
studied earlier in the literature 
\cite{ST,HKLR,BGIO,DFF,ABCDFFM} but our purpose is to incorporate 
them into the set-up discussed in this section.

\section{Hypermultiplets with gauged target-space isometries}

The above Lagrangian and transformation rules are subject to two 
classes of equivalence transformations associated with the target 
space. One class consists of the target-space diffeomorphisms associated with 
$\phi\to \phi^{\prime}(\phi)$. The other refers to reparametrizations of 
the fermion `frame' of the form $\zeta^\a\to  S^\a{}_\b(\phi)\, \z^\b$, 
and similar redefinitions of other quantities carrying indices 
$\a$ or $\bar\a$. For example, the fermionic metric transforms as 
$G_{\bar\a\b} \to [\bar S^{-1}]^{\bar\g}{}_{\bar\a}\,
[S^{-1}]^\d{}_{\b}\, G_{\bar\g\d}$. 
Under these rotations the quantities 
$\G_A{}^{\!\a}{}_\b$ play the role of connections. 

The above transformations do not constitute invariances of the 
theory. This is only the case when the metric $g_{AB}$ and 
and the Sp($n$)$\times$Sp(1) one-form $V^\a_i$ (and thus the
related geometric quantities) are left invariant under 
(a subset of) them. To see how this works, let us consider the
scalar fields transforming under a certain  
isometry (sub)group G characterized by a number of Killing vectors 
$k^A_I(\phi)$, with parameters $\theta^I$. 
Hence under infinitesimal transformations, 
\begin{equation}
\delta_{\rm G} \phi^A=g\,\theta^Ik_I^A(\phi)\ ,
\end{equation}
where $g$ is the coupling constant and the $k^A_I(\phi)$ satisfy 
the Killing equation
\begin{equation}
D_Ak_{IB} + D_B k_{IA}=0\,. \label{killing-eq}
\end{equation} 
The quantities such as $V^\a_{Ai}$ that carry Sp($n$) indices are
only required to be 
invariant under isometries up to fermionic equivalence 
transformations. Thus $-g(
k^B_I \,\pa_B V^\a_{Ai} + \partial_A k_I^B\, V^\a_{Bi})$ must be
cancelled by a suitable   
infinitesimal rotation on the index $\a$. 
Here we make the important assumption that the effect of  
the diffeomorphism is entirely compensated by a rotation that
affects the indices $\a$. In principle, one 
can also allow a compensating Sp(1) transformation acting on the indices
$i,j,\ldots$. However, we will not do this here, as this would
imply that the isometry group would neither commute with Sp(1)
nor with supersymmetry. Instead we return to this option
in the next section.  

Let us parametrize the compensating transformation
acting on the Sp($n$) indices by 
$\delta_{\rm G} \z^\a= g[{t_I} - k_I^A\, \Gamma_A]^{\a}{}_{\!\b} \,\z^\b$, 
where the ($\phi$-dependent) matrices $t_I(\phi)$ remain to be
determined, 
\be
-k^B_I \,\pa_B V_{A\,i}^\a - \pa_A k^B_I \,V_{B\,i}^\a + 
(t_I -k^B_I\,\G_B)^{\a}{}_\b \, V_{A\,i}^{\!\b} =0\,. 
  \label{invariant-bein}
\ee
Obviously similar equations apply to the other geometric
quantities, but as those are not independent we do not need to
consider them. 

Subsequently we derive the main consequences of the two
equations \eqn{killing-eq} and \eqn{invariant-bein}. First of
all the isometries must constitute an algebra with certain structure
constants. This is expressed by
\begin{equation}
k^B_I\partial_Bk^A_J-k^B_J\partial_Bk^A_I =-f_{IJ}{}^K\, k^A_K\ ,
\label{killingclosure}
\end{equation} 
where our definitions are such that the gauge fields that are
needed 
once the $\theta^I$ become spacetime dependent, transform according to  
$\delta_{\rm G}W^I_\m=\partial_\mu\theta^I- gf_{JK}{}^I\,W^J_\mu\,
\theta^K$.
The Killing equation implies the following property 
\begin{equation}
D_{A}D_{B} k_{IC}  = - R_{BCAD}\,  k^{\,D}_I  \,. 
\label{symmetric}
\end{equation}
Then, using the covariant constancy of 
$V_A$, we find from \eqn{invariant-bein},
\be
(t_I)^{\a}{}_{\!\b} = \ft12 V_{Ai}^{\a} \,
\bar\gamma^{Bi}_{\b}\, D_Bk_I^{\,A}\,. 
\ee
Target-space scalars satisfy algebraic identities, e.g.,
\be
(\bar t_I)^{\bar\g}{}_{\!\bar\a} \, G_{\bar\g\b} 
+(t_I)^{\g}{}_{\!\b} \, 
G_{\bar\a\g}= (t_I)^{\bar\g}{}_{\![\bar\a} \, \Omega_{\bar\b]\bar\g} = 0\,,
\ee
which shows that the field-dependent matrices $t_I$ take 
values in $sp(n)$. An explicit calculation, making use 
of the equations \eqn{killing-eq} and \eqn{symmetric}, shows that
\begin{equation}
D_A t_I{}^{\!\a}{}_{\!\b}  =  k^{\,B}_I\, 
R_{AB}{}^{\!\a}{}_{\!\b} \,, \label{t-der}
\end{equation}
for any infinitesimal isometry. From the group property of the
isometries it follows that 
the matrices $t_I$ satisfy the commutation relation 
\be
[\,t_I ,\,t_J\,]^\a{}_{\!\b}   = f_{IJ}{}^K\, 
(t_K)^\a{}_{\!\b}  + k^A_I\,k^B_J\, 
R_{AB}{}^{\!\a}{}_{\!\b} \,.  \label{t-comm}
\ee
The apparent lack of closure represented by the presence of the
infinitesimal Sp($n$) holonomy transformation is related to the fact that the
coordinates $\phi^A$ on which the matrices depend, transform
under the action of the group. One can show that this result is
consistent with the Jacobi identity. 

Furthermore we derive from 
\eqn{invariant-bein} that the complex structures $J_{AB}^{ij}$ are 
invariant under the isometries, 
\be
k^C_I \pa_C J_{AB}^{ij} - 2 \pa_{[A}k^C_I \,J_{B]C}^{ij}  =
0\,. \label{triholo} 
\ee
This means that the isometries are {\it tri-holomorphic}. From
\eqn{triholo} one shows that $\pa_A(J^{ij}_{BC}\, k^C_I )
-\pa_B(J^{ij}_{AC}\, k^C_I )=0$, 
so that, locally, one can associate three Killing potentials (or
moment maps) $P^{ij}_I$ to every Killing vector, according to
\be
\pa_A P_I^{ij}  = J_{AB}^{ij} \,k^{\,B}_I \,.
\ee
Observe that this condition determines the moment maps up to a 
constant. Up to constants one can also derive the 
equivariance condition,
\begin{equation}
J^{ij}_{AB}\,k^A_Ik^B_J=-f_{IJ}{}^{\!K}\,P^{ij}_K\ , \label{equivariance}
\end{equation}
which implies that the moment maps transform covariantly under 
the isometries,
\be
\d_{\rm G}P_I^{ij} = \theta^J\,k_J^A\,\pa_A P^{ij}_I  = - 
f_{JI}{}^{\!K} \,P^{ij}_K\,\theta^J \, . \label{moment-isometry}
\ee

Summarizing, the invariance group of the isometries acts as follows,
\begin{equation}
\d_{\rm G}\phi = g\,\theta^I\,k_I^{\,A}\,,\qquad \delta_{\rm G} 
\z^\a=g\, 
(\theta^I{t_I})^{\a}{}_{\!\b}\,\z^\b - \d_{\rm G}\phi^A  
\Gamma_A{}^{\!\a}{}_{\!\b} \,\z^\b \,.\label{fermgaugetr}
\end{equation}
When the parameters of these isometries become spacetime 
dependent we introduce corresponding gauge fields and fully 
covariant derivatives, 
\bea
D_\mu \phi^A = \partial_\m \phi^A - g W^I_\m \,k_I^A \,
, \quad
D_\mu\z^\a =\partial_\mu \z^\a+\pa_\mu\phi^A\,
{\G_A}^{\!\a}{}_{\!\b}  
\z^\b -gW_\m{}^{\!\!\a}{}_{\!\!\b}\z^\b\, , \;
\eea
where $W_\m{}^{\!\a} {}_{\!\b} =W^I_\m\,(t_I)^{\a}{}_{\!\b}$. The 
covariance of $D_\m\zeta^\a$ depends crucially on \eqn{t-der} and 
\eqn{t-comm}. The gauge
fields $W^I_\mu$ are accompanied by complex scalars $X^I$, spinors
$\Omega_i^I$ and auxiliary fields $Y_{ij}^I$, constituting off-shell
$N=2$ vector multiplets. For our notation of vector multiplets,
the reader may consult \cite{DDKV}. 

The minimal coupling to the gauge fields requires extra terms in
the supersymmetry transformation rules for the hypermultiplet
spinors as well as in the Lagrangian, in order to regain $N=2$
supersymmetry. The extra terms in the transformation rules are
\be
\delta^\prime _{\scriptscriptstyle\rm Q}\z^\a = 
2gX^Ik_I^AV^\a_{Ai}\,\varepsilon^{ij}\e_j\,,\qquad  
\delta^\prime_{\scriptscriptstyle\rm Q}\z^{\bar \a}= 2g{\bar X}^Ik_I^A{\bar V}^{{\bar \a} 
i}_{A}\,\varepsilon_{ij}\e^j\ . \label{susyferm} 
\ee
These terms can be conveniently derived by imposing the commutator of
two supersymmetry transformations on the scalars, as this
commutator should yield the correct field-dependent gauge
transformation.  

We distinguish three additional couplings to the Lagrangian. The
first one is quadratic in the hypermultiplet spinors and reads
\begin{eqnarray}
\lagr_g^{(1)} =g{\bar X^I}{\bar \g}^{Ai}_\a\e_{ij}{\bar
\g}^{Bj}_\b\,D_Bk_{AI}\,{\bar \z}^\a\z^\b + \mbox{h.c.}
=2g{\bar X}^I{t_I}^{\!\gamma}{}_{\!\a}\,\O_{\b\gamma}\,{\bar \z}^\a\z^\b+
\mbox{h.c.} \ .\label{S1}
\end{eqnarray}
The second one is proportional to the vector multiplet spinor
$\O^I$ and takes the form
\begin{equation}
\lagr_g^{(2)} =-2gk^A_IV_{Ai}^\a\O_{\a\b}\,{\bar \z}^\b\O^{Ii} +
\mbox{h.c.} =2gk^A_I{\bar \g}^i_{A\a}\e_{ij}\,{\bar
\z}^\a\O^{Ij}  + \mbox{h.c.} \ .\label{S2}
\end{equation}
Finally there is a potential given by
\begin{equation}
\lagr_{g}^{\rm scalar} =-2g^2k^A_Ik^B_J\,g_{AB}\,X^I{\bar X^J} + g \,
P^{ij}_I\, Y^I_{ij}\ ,\label{pot}
\end{equation}
where $P^{ij}_I$ is the triplet of moment maps on the
hyper-K\"ahler space. These terms were determined both from imposing
the supersymmetry algebra and from the invariance of the action. 
To prove \eqn{pot}, one has to make use of the equivariance
condition \eqn{equivariance}. Actually, gauge invariance, which 
is prerequisite to supersymmetry, already depends on 
\eqn{moment-isometry}.

\section{Superconformally invariant hypermultiplets}
In this last section we determine the restrictions
from superconformal invariance on the hypermultiplets by 
evaluating some of the couplings to the fields of $N=2$ 
conformal supergravity. At this stage we have only a
modest goal, namely to determine the restrictions on the
hyper-K\"ahler geometry that arise from requiring invariance
under rigid superconformal transformations.  This is the
situation that arises when freezing all the fields of conformal
supergravity to zero in a flat spacetime metric. In that case the 
superconformal transformations acquire an explicit dependence on 
the spacetime coordinates.

We start by implementing the $N=2$ superconformal algebra \cite{DVV} 
on the hypermultiplet fields. We assume that the scalars are invariant 
under special conformal and special supersymmetry 
transformations, but they transform under $Q$-supersymmetry and 
under the additional   
bosonic symmetries of the superconformal algebra, namely chiral 
[SU(2)$\times$U(1)]$_{\rm R}$ and dilatations denoted by $D$. At 
this point we do not assume  
that these transformations are symmetries of the action and we 
simply parametrize them as follows,
\be
\d\phi^A = \thd\, \kild{A}(\phi) + \theta_{\scriptscriptstyle\rm U(1)}\, 
k^A_{\scriptscriptstyle\rm U(1)}(\phi)  
+(\theta_{\scriptscriptstyle\rm SU(2)})^i{}_k \,\varepsilon^{jk} 
k^A_{ij}(\phi) \,,
\ee
where the $k^A$ are left arbitrary. Note that  $k^A_{ij}(\phi)$ 
is assigned to the same symmetric pseudoreal representation of 
SU(2) as the complex structures, while 
$\theta_{\scriptscriptstyle\rm SU(2)}$ is antihermitean and 
traceless.  

An important difference with the situation described in the 
previous section,  is that in the conformal superalgebra 
the dilatations and chiral transformations do not appear in the 
commutator of two $Q$-supersymmetries, but in the commutator of a 
$Q$- and an $S$-supersymmetry. To evaluate the $S$-supersymmetry 
variation of the fermions, we use that $\d_{\scriptscriptstyle\rm 
S}\phi^A = \d_{\scriptscriptstyle\rm K} \zeta^\a=0$   
and covariantize the derivative in the fermionic transformations 
with respect to dilatations. Subsequently we impose the 
commutator, $[\,\delta_{\scriptscriptstyle\rm 
K}(\Lambda_{\scriptscriptstyle\rm K}),
\delta_{\scriptscriptstyle\rm Q}(\epsilon)\,] =  
-\delta_{\scriptscriptstyle\rm S}(\,\rlap/\!\L_K \epsilon)$ 
on the spinors. This expresses the $S$-supersymmetry variations 
in terms of $\kild{A}$,
\be 
  \delta_{\scriptscriptstyle\rm S}(\eta)\,\z^{\a} = {V}^{\a}_{i\,A} \, 
\kild{A}\,\eta^i\,,\qquad 
  \delta_{\scriptscriptstyle\rm S}(\eta)\,\z^{\bar\a} = 
{\bar{V}}^{i\,\bar\a}_A \,  \kild{A}\,\eta_i\,.
  \label{eq:dSTwo}
\ee
With this result we first evaluate the commutator of an $S$- and a 
$Q$-supersymmetry transformation on the scalars. This yields
\be 
{}[\,\d_{\scriptscriptstyle\rm S}(\eta), 
\d_{\scriptscriptstyle\rm Q}(\e)\,] \,\phi^A = (\bar \e^i\eta_i   
+\bar\e_i\eta^i) \, \kild{A} + 2J_{ik}{}^{\!\!A}{}_{\!B} \,
\varepsilon^{kj}\,
(\bar \e^i\eta_j -\bar\e_j\eta^i) \,\kild{B}\,. 
\ee
This result can be confronted with the universal result from 
$N=2$ conformal supergravity, which reads 
\bea
{}  [\,\delta_{\scriptscriptstyle\rm S}(\eta), 
\delta_{\scriptscriptstyle\rm Q}(\epsilon)\,] &=&  
    \delta_{\scriptscriptstyle\rm M} 
(2\bar{\eta}^i\s^{ab}\epsilon_i + {\hbox{h.c.}}) 
    +\delta_{\scriptscriptstyle\rm D} (\bar{\eta}_i \epsilon^i + 
{\hbox{h.c.}})\nonumber\\ 
  & &+\delta_{\scriptscriptstyle\rm U(1)} (i\bar{\eta}_i 
\epsilon^i +{\hbox{h.c.}}) 
    +\delta_{\scriptscriptstyle\rm SU(2)} (-2  \bar{\eta}^i \epsilon_j 
-{\hbox{h.c.} \, ; \, {\rm traceless})}\,.   
  \label{eq:SQComm}
\eea
Comparison thus shows that $k^A_{\scriptscriptstyle\rm U(1)}$ 
vanishes and that the SU(2) vectors satisfy
\be
k^A_{ij} =  J_{ij}{}^{\!A}{}_{\!B}\, \kild{B}\,.  
\label{su(2)-vector}
\ee
Now we proceed to impose the same commutator on the fermions, 
where on the right-hand side we find a Lorentz transformation, a 
U(1) transformation and a dilatation, iff we assume the 
following condition on $\kild{A}$,
\be
D_A\kild{B} = \d^B_A\,.  \label{homo-vector}
\ee
This condition suffices to show that the kinetic term of the scalars 
is scale invariant, provided one includes a spacetime metric or, in 
flat spacetime, include corresponding scale transformations of 
the spacetime coordinates. Nevertheless, observe that $\kild{A}$ 
is {\it not} a Killing vector of the hyper-K\"ahler space, but 
a special example of a conformal homothetic Killing vector (we 
thank G. Gibbons for an illuminating discussion regarding such 
vectors). An immediate consequence of \eqn{homo-vector} is that 
$\kild{A}$ can (locally) be expressed in terms of a potential 
$\chi_{\scriptscriptstyle\rm D}$, according to 
$k_{{\scriptscriptstyle\rm D}\,A} 
=\pa_A\chi_{\scriptscriptstyle\rm D}$. Another consequence is that  
the SU(2) vectors $k^A_{ij}$, as expressed by \eqn{su(2)-vector}, 
are themselves Killing vectors,  
because their derivative is proportional to the corresponding 
antisymmetric  complex structure,
\be
D_Ak_B^{ij} = - J_{AB}^{ij}\,. 
\ee
Therefore, the bosonic action is also invariant under these 
SU(2) transformations. 

From the $[\d_{\scriptscriptstyle\rm S},\d_{\scriptscriptstyle\rm 
Q}]$ commutator we also find the fermionic 
transformation rules under the chiral transformations and the 
dilatations,
\be
\begin{array}{l} 
  \delta_{\scriptscriptstyle\rm SU(2)}\,\zeta^{\a} + 
\delta_{\scriptscriptstyle\rm SU(2)}\phi^A\,
    \Gamma_A{}^{\!\a}{}_{\!\b}\zeta^\b = 0\, ,\\
 \delta_{\scriptscriptstyle\rm U(1)}\zeta^\a + 
\delta_{\scriptscriptstyle\rm U(1)}\phi^A \, 
\Gamma_A{}^{\!\a}{}_{\!\b}\zeta^\b
= -\ft12i\,\theta_{\scriptscriptstyle\rm U(1)}\zeta^\a\,,  
\end{array}
\qquad
\begin{array}{l}
  \delta_{\scriptscriptstyle\rm D}\zeta^\a + 
\delta_{\scriptscriptstyle\rm D}\phi^A\,
    \Gamma_A{}^{\!\a}{}_{\!\b}\zeta^\b = \ft32\,
\theta_{\scriptscriptstyle\rm D}\zeta^\a\,.
\end{array}
\ee
Note that the U(1) transformation further simplifies because 
$\delta_{\scriptscriptstyle\rm U(1)}\phi^A=0$. 

To establish that the model as a whole is now invariant under the 
superconformal transformations it remains to be shown that the 
tensor $V^\a_{Ai}$ is invariant under the diffeomorphisms 
generated by $k_{ij}^A$, $k_{\scriptscriptstyle\rm U(1)}^A$ and 
$k_{\scriptscriptstyle\rm D}^A$ up to compensating transformations that act on 
the Sp($n$)$\times$Sp(1) indices in accordance with the 
transformations of the $\zeta^\a$ given above and the symmetry 
assignments of the supersymmetry parameters $\e^i$. To emphasize 
the systematics we ignore the fact that $k_{\scriptscriptstyle\rm 
U(1)}^A$ actually vanishes and we write 
\bea
- k_{kl}^B\, \pa_B V^\a_{Ai} - \pa_Ak^B_{kl}\,  V^\a_{Bi} -  
k^B_{kl}\,\Gamma_B{}^{\!\a}{}_{\!\b} V^\b_{Ai} + 
[- \d_{(k}^j\varepsilon^{~}_{l)i}]\, 
V^\a_{Aj} &=& 0\,,
\nonumber\\
- k^B_{\scriptscriptstyle\rm U(1)} \pa_B V^\a_{Ai} - 
\pa_Ak^B_{\scriptscriptstyle\rm U(1)} \, V^\a_{Bi} +
[- \ft12 i\,\d^\a_\b- k^B_{\scriptscriptstyle\rm U(1)} \,
\Gamma_B{}^{\!\a}{}_{\!\b}]  V^\b_{Ai} +  [\ft12 i\,\d^j_i]\, 
V^\a_{Aj} &=& 0\,, 
\nonumber\\
- k^B_{\scriptscriptstyle\rm D} \,\pa_B V^\a_{Ai} - 
\pa_Ak^B_{\scriptscriptstyle\rm D} \,V^\a_{Bi} +
[\ft32 \d^\a_\b-  \kild{B}\,\Gamma_B{}^{\!\a}{}_{\!\b} ]\,  
V^\b_{Ai} + [- \ft 12\, \d^j_i]\, V^\a_{Aj} &=& 0\,. \;\;
\label{comp-sc}
\eea
In these equations the first two terms 
on the left-hand side represent the effect of the isometry, the 
third terms represent a uniform scale and chiral U(1) 
transformation on the indices associated with the Sp($n$) tangent 
space, and the last terms represent an SU(2), a U(1) and a scale 
transformation, respectively, on the indices associated with the 
Sp(1) target space. Eq. \eqn{comp-sc} should be regarded as a 
direct extension of \eqn{invariant-bein}. 

We close with a few comments. First of all, the SU(2) isometries 
induce a rotation on the complex structures, 
\be
k_{kl}^C\,\pa^{~}_C J^{ij}_{AB} - 2\pa_{[A}k^C_{kl}\,
J_{B]C}^{ij} = -2 J_{klC[A} \,J^{ijC}_{\,B]}= 2 \d_{(k}^{(i} \, 
\varepsilon^{~}_{l)m}\,J ^{j)m}_{AB}\,,
\ee
as should be expected. Secondly, one can verify that the 
isometries discussed in section 3 commute with the scale and 
chiral transformations, provided that $k_I^A = \kild{B}\,D_B 
k^A_I$. This condition is also required for the scale invariance 
of the full action. Observe also that $\kild{A}$ satisfies \eqn{symmetric}, 
in spite of the fact that it is not a Killing vector. Finally, it 
is straightforward to write down actions for the vector 
multiplets that are invariant under rigid $N=2$ superconformal 
transformations. Those are based on a holomorphic 
function that is homogeneous of degree two.

\vspace{8mm}
\noindent
{\bf Acknowledgement}\\
S.~V. thanks PPARC for financial support and the Institute for 
Theoretical Physics in Utrecht for its hospitality.

%


\begin{thebibliography}{99}
%
\bibitem{sconf}
J. Maldacena, {\it The large-$N$ 
limit of superconformal field theories and supergravity} ({\tt 
hep-th/9711200}). 
%
\bibitem{DDKV}
J. De Jaegher, B. de Wit,
B. Kleijn and S. Vandoren, Nucl. Phys. {\bf B514} (1998) 553 ({\tt 
hep-th/9707262}). 
%
\bibitem{BagWit}
J.~Bagger and 
E.~Witten, Nucl.~Phys. {\bf B222} (1983) 1.
%
\bibitem{HKLR}
C.M. Hull, A. Karlhede, U. 
Lindstr\"om and M. Ro\v{c}ek, Nucl. Phys. {\bf B266} (1986) 1.
%
\bibitem{ST}
G. Sierra and P.K. 
Townsend, Nucl. Phys. {\bf B233} (1984) 289. 
%
\bibitem{BGIO}
J.A. Bagger, A.S. 
Galperin, E.A. Ivanov and V.I. Ogievetsky, Nucl. Phys. {\bf B303} 
(1988) 522.   
%
\bibitem{DFF}
R.~D'Auria, S.~Ferrara and 
P.~Fr\'e, Nucl.~Phys. {\bf B359} (1991) 705.
%
\bibitem{ABCDFFM}
L. Andrianopoli, 
M. Bertolini, A. Ceresole, R. D'Auria, S.  
Ferrara, P. Fr\'e and T. Magri, J. Geom. Phys. {\bf 23} (1997) 
111 ({\tt hep-th/9605032}). 
%
\bibitem{DVV} 
B. de Wit, J.W. van Holten 
and A. Van Proeyen, Nucl. Phys. {\bf B167} (1980) 186.
%
%
\end{thebibliography}
\end{document}